\documentclass[12pt]{article}
\usepackage[cp1251]{inputenc}
\usepackage[english]{babel}

\usepackage{indentfirst}
\usepackage{graphicx}

\newcommand{\di}{\displaystyle}

\fussy
\begin{document}

{\bf Motion of Stars in Layered Inhomogeneous Elliptical Galaxies}
\vskip 2ex
S. A. Gasanov 
\vskip 2ex
Sternberg Astronomical Institute, Moscow State University, Moscow, Russia
\vskip 2ex
e-mail: gasanovsa57@gmail.com
\vskip 2ex

{\bf Abstract.} The problem of the spatial motion of a passively gravitating body (PGB) in the gravitational field
of a layered inhomogeneous elliptical galaxy (LIEG) is considered on the basis of the previously developed
model. It is assumed that a LIEG consists of baryonic mass (BM) and dark matter (DM), which have differ-
ent laws of density distribution. A star or the center of mass of a globular cluster is taken as the PGB, the
motion of which considers the BM and DM attraction. To obtain accurate results, the BM and DM attraction
potentials are not expanded in a series, but their exact expressions are taken. An analogue of the Jacobi inte-
gral is found, the region of the possible motion of the PGB is determined, and the zero-velocity surfaces are
constructed. The stationary solutions (libration points) are found to be stable in the sense of Lyapunov. The
results are applied to the elliptical galaxies NGC 4472 (M 49), NGC 4697, and NGC 4374 (M 84).

\vskip 2ex


\vskip 4ex

1. INTRODUCTION
\vskip 2ex

The problem of the spatial motion of a passively
gravitating body (PGB) in the gravitational field of an
elliptical galaxy (EG) according to models 1 and 2 is
considered in [1, 2]. A similar problem of the PGB
motion inside (near) a globular cluster (GC) belonging to an EG is studied in [3].
The results obtained in [1–3] are applied to model
elliptical galaxies with parameters that exactly coincide with the parameters of the elliptical galaxies NGC
4472 (M 49), NGC 4697, and NGC 4374 (M 84);
these results are presented in the form of figures and
atable.
Three new EG models (Models 3, 4, and 5) are
considered in [4]. Of greatest interest is Model 5,
according to which the EG with a halo (option 1) or
without it (option 2) is represented as an inhomogeneous ellipsoid of revolution, i.e., an elongated 
spheroid that consists of BM and DM. Such a spheroid was
chosen as a model of a triaxial EG because its dynamic
properties are very close to those of a triaxial ellipsoid
[5]. In addition, the fulfillment of the potential matching conditions should not be considered in Model 5,
since there is no interface between the BM and DM in
the galaxy. The results obtained in [4] are applied to
sixty EGs and are presented in the form of tables for
ten of them.
The models mentioned above are intended for
solving problems of celestial mechanics and partially
astrophysics. Another attempt to study the impact of
the DM on the kinematics and dynamics of PGB is
made within the context of these models. These models cannot claim to provide a complete coverage of the
DM problem as a whole. Moreover, according to some
authors [6], the bulk of the DM lies outside the luminous part of an elliptical galaxy, while others believe
[7, 8] that the DM content in the inner regions of an
EG is comparable to the BM content.
This study considers the problem of the spatial
motion of a PGB in the gravitational field of an LIEG
that has the shape of an elongated spheroid. Model 5 is
used as the basis. An analogue of the Jacobi integral is
found, the region of the possible motion of the PGB is
determined, and the zero-velocity surfaces are constructed. The stationary solutions (libration points)
are found to be stable in the sense of Lyapunov.
The question of equilibrium and stability of the
dynamical system studied in Model 5 will be considered separately in another paper of the author.

\vskip 4ex

2. STATEMENT OF THE PROBLEM. 
EQUATIONS OF MOTION

\vskip 2ex

Let us consider the problem of the spatial motion of
a PGB in the LIEG gravitational field according to
Model 5 in the coordinate system $OXYZ$. The conditional boundaries of such a galaxy are determined by
the $D_{25}$ and $R_{25}$ values [9].  $OXYZ$ is a coordinate system with the origin at the center of the EG rotating at
a constant angular velocity  $\Omega $ around the polar axis $OZ$
and with axes directed along the corresponding principal axes of the EG. The rectangular coordinates 
$(x, y, z)$ of the PGB in this coordinate system are determined from the system of equations [10]
$$
 \frac{d^2x}{dt^2} - 2\Omega\, \frac{dy}{dt} = \frac {\partial U}{\partial x}, \quad
\frac{d^2y}{dt^2} + 2\Omega\,\frac {dx}{dt} = \frac {\partial U }{\partial y}, \quad
 \frac{d^2z}{dt^2}  =  \frac{\partial U}{ \partial z}
\eqno (1) $$
Here, the force function  $U = U\,(x, y, z)$  is defined by
the equality
$$
  U  = \frac{\Omega^2}{2}\,(x^2 + y^2) + V, \quad V = U^* + U_G,
 \eqno (2)  $$
where the first term is the centrifugal force potential,  $V = V\,(x, y, z)$
 is the potential of the force of attraction,
and function $U$  can be considered the potential of
gravity.  $U^* = U^*(x, y, z)$ and $U_G = U_G(x, y, z)$   are the
potentials of the LIEG’s BM and DM, respectively,
the explicit form of which is given in the following sections.

For an inhomogeneous EG to exist as a figure of
equilibrium, the necessary condition of the Poincare
inequality for the angular velocity of rotation [11] must
be satisfied:
$$
  \Omega^2 \le 2 \pi G  \bar \rho, \quad ( \Omega^2 \le  \pi G   \rho_0, \quad   \Omega^2 \le 0.4 \pi G  \bar \rho)
\eqno (3)  $$
Here, $G$ is the gravitational constant, and $\bar \rho$ is the
average density of an inhomogeneous elliptical galaxy.
The fulfillment of the Poincare inequality guarantees
that the total force of gravity is oriented inward and the
pressure is non-negative. The stricter Crudeli and
Kondratyev inequalities [12] are indicated in parentheses. In the Crudeli inequality,  $\rho_0$ is the density in
the center of the galaxy and it decreases from the center to the periphery. In addition, the direction of gravity is not involved.

Obviously, the system of equations (1) allows an
analogue of the Jacobi integral in the form [1, 2, 10]
$$
     \left(\frac{dx}{dt} \right)^2 + \left(\frac{dy}{dt} \right)^2 + \left(\frac{dz}{dt} \right)^2 = 2U - 2C, 
     \quad C = \mbox{const},
$$
from which zero-velocity surfaces and the region of
possible motion of the PGB are easily obtained:
$$
     U  = C, \qquad   U  \ge C 
 $$
respectively, where $C$ is the analogue of the Jacobi constant.

\vskip 4 ex

3. BARYONIC MASS POTENTIAL OF A LAYERED INHOMOGENEOUS  ELLIPTICAL GALAXY

\vskip 2ex

We will assume that the LIEG has the form of an
inhomogeneous elongated spheroid bounded by a
spheroidal surface,
$$
   \frac{x^2}{a^2} +  \frac{w^2}{c^2} = m^2, \quad  w^2 = y^2 + z^2,  \qquad (a \ge b = c, \quad  0 \le m \le 1),
\eqno (4) $$
where the value of the family parameter $m = 0$  corresponds to the center of the LIEG, and $m = 1$ to its
outer border. The distribution laws of density $\rho\,(m)$ and
surface brightness $I\,(m)$ of the BM are described by the
expressions [3, 13, 14]
$$
     \rho\,(m) = \frac{\rho_0}{\left(1 + \di \beta m^2\right)^{3/2}}, \quad 
         I\,(m) = \frac{I_0}{1 + \di \beta m^2}
\eqno (5) $$
respectively. Here, $\rho_0$ is the density of the center (core)
of an elliptical galaxy, $m$ is the parameter of the family
of ellipsoidal surfaces (5) that comprise its luminous
part, and parameter $\beta \gg 1$ is selected separately for
each EG [13] and is found by aligning the photometry
data [12, 13]. $I_0$ is the central surface brightness.
The profile in the form (5) will be called "astrophysical" in accordance with [12]; it is consistent with
modern concepts of the EG structure [13, 14].

The attraction potential of the LIEG BM with density $\rho\,(m)$   to the outer point $P = P\,(x, y, z)$ and 
its derivatives with respect to the coordinate axes are determined by the equalities [10, 12]
$$
        U^*(P) =   \pi G ac^2 \int \limits_{\lambda}^{\infty} \frac{\delta\,(m^2(u))}{\Delta\,(u)}\, du,  \quad    
      \frac{\partial  U^*(P)}{\partial R} =  -\,2 \pi G a c^2 \,R\,\int \limits_{\lambda}^{\infty}  \frac{\rho\,(m^2(u))\, du}{(A^2 + u)\,\Delta\,(u)},  
\eqno (6) $$
where $R$ is one of the coordinates $\{x, y, z\}$, and $A^2 = \{a^2, c^2\}$. If the BM profile is determined by
equality (5),
$$
  \delta\,(m^2(u)) =  \int \limits_{m^2(u)}^1  \rho\,(v) dv^2 =
  \frac{2\rho_0}{\beta}\,\left[-\,\frac{1}{\di \sqrt{1 + \beta}} + \frac{1}{\di \sqrt{1 + \beta\, m^2(u)}} \right],
$$   
$$
      m^2(u) = \frac{x^2}{a^2 + u} + \frac{w^2}{c^2 + u},   \quad \Delta^2(u) = (a^2 + u)(c^2 + u)^2,  
      \quad m^2(0) = \frac{x^2}{a^2} + \frac{w^2}{c^2} > 1,  
\eqno (7) $$
and the fulfillment of the condition $m^2(0) > 1$ for the
coordinates of the outer point $P$ of the luminous part
of the EG is mandatory. Parameter $\lambda$ is the positive
root of the quadratic equation  $m^2(\lambda) = 1$:
$$
   \lambda^2 + p \lambda + q = 0, \quad \lambda = \frac{1}{2}\,\left(-\,p + \sqrt{\delta}\right), 
   \quad \delta = p^2 - 4q > 0,
$$
where 
$$
     p = a^2 + c^2 - x^2 - w^2, \quad  
  q = a^2c^2\left(1 - \frac{x^2}{a^2} - \frac{w^2}{c^2}\right) =  a^2c^2 \left[ 1 - m^2(0) \right] < 0
$$
By virtue of (7), for $p > 0$, we have
$$
  c^2m^2(0) < x^2 + w^2 < a^2m^2(0), \quad  a^2 + c^2 - a^2 m^2(0) < p <  a^2 +  c^2 - c^2m^2(0) 
$$
Obviously, the potential $U^*$  is a function of parameter
 $\lambda$, which, in turn, depends on the coordinates of the
outer point $P = P\,(x, y, z)$. After calculating the integral in (6) for the outer potential, we obtain
$$
   U^*(P) \equiv U^*(\lambda) = \frac{2 \pi G \rho_0 ac^2}{\beta}\,
  \left[ -\,\frac{U_1(\lambda)}{\di \sqrt{1 + \beta}} + U_2(\lambda)\right],  
\eqno (8) $$
where 
$$
   U_1(\lambda) =    \int \limits_{\lambda}^{\infty} \frac{du}{\di (c^2 + u)\,\sqrt{a^2 + u}} =    \frac{\ln \varphi\,(\lambda)}{\di \sqrt{a^2 - c^2}}, \quad 
   \varphi\,(\lambda) =  \frac{\di \sqrt{a^2 + \lambda} + \sqrt{a^2 - c^2}}{\di \sqrt{a^2 + \lambda} - \sqrt{a^2 - c^2}},
$$
$$
   U_2(\lambda) =   \int \limits_{\lambda}^{\infty} \frac{du}{\di \di (c^2 + u)\,\sqrt{a^2 + u}\, \sqrt{1 + \beta\, m^2(u)}}
 =  \int \limits_{\lambda}^{\infty}  \frac{du}{\di \sqrt{c^2 + u}\,\sqrt{(u - v_1)(u - v_2)}} =  \frac{2F\,(\alpha,  n)}{\di \sqrt{-\, c^2 -  v_2}} 
$$
Here,  
$$
 v_1 = \frac{-\,p_1 + \sqrt{\delta_1}}{2} < 0, \quad    v_2 = \frac{-\,p_1 - \sqrt{\delta_1}}{2} < 0,  \quad \delta_1 = p_1^2 - 4q_1 > 0, 
$$
$$
   p_1 = a^2 + c^2 + \beta\,( x^2 +  w^2) > 0, \quad   q_1 =  a^2c^2 \left(1 + \beta\,\frac{x^2}{a^2} +    \beta\,\frac{w^2}{c^2} \right)  > 0,
$$
and
$$
 v_1 + v_2 = -\,p_1 < 0, \quad v_1v_2 = q_1 > 0, \quad c^2 + v_2 <  c^2 + v_1 < 0,
  \quad  (0 > v_1 > v_2)
$$
Argument $\alpha$ and modulus $n$  of an incomplete elliptic
integral of the first kind  $F\,(\alpha,  n)$ are
$$
   \alpha = \di \sqrt{\frac{-\,c^2 - v_2}{\lambda - v_2}}, \quad 
  n = \sqrt{\di \frac{v_1 - v_2}{-c^2 - v_2}} < 1
$$
At the origin of coordinates (in the center of the galaxy), it is obvious that the value of 
the potential $U^*(\lambda)$ considering the equalities
$$
  x = 0, \quad w = 0, \quad m = 0, \quad m^2(u) = 0, \quad \lambda = 0, \quad \rho\,(m = 0) = \rho_0, \quad 
   U_2(\lambda) \equiv  U_1(\lambda)  
$$ 
will be
$$
    U_0^* =  \frac{2 \pi G \rho_0 ac^2}{\beta}\,\frac{\ln \varphi\,(0)}{\di \sqrt{a^2 - c^2}}\,
    \left(1 - \frac{1}{\di \sqrt{1 + \beta}}\right), \quad 
    \varphi\,(0) =  \varphi\,(\lambda = 0), 
$$
and function $\varphi\,(\lambda)$ is defined above.

The last equality can also be obtained from expression (8) of the potential  $U^*(\lambda)$ setting 
 $U_2(\lambda) = U_1(\lambda)$ and  $\lambda = 0$.

Setting $z = 0,  x \neq 0$, and $w = y \neq 0$ in the corresponding expressions, we find the parameter $\lambda$, 
the roots $v_1$ and  $v_2$, then the values of the functions 
$U_1(\lambda)$ and $U_2(\lambda)$. Further, we obtain the expression for the
potential $U^*(\lambda)$ in the plane $Oxy$. Similarly, at $y = 0$, $x \neq 0$, $w = z \neq 0$,
 we find the expression for potential $U^*(\lambda)$  in the plane $OXZ$.
 
Further, in the plane $OYZ$ we have
$$
  x = 0, \quad    m^2(u) = \frac{w^2}{c^2 + u}, \quad    \lambda = w^2 - c^2, 
  \quad v_1 = -\,a^2, \quad v_2 = -\,c^2 - \beta w^2
$$
The potential $U^*(\lambda)$  will then be described by exactly
the same equality as (8) with the only difference that
the parameter $\lambda$ and function $U_2(\lambda)$ are defined differently:
$$
  U_2(\lambda) = \frac{2 F\,(\alpha_0,  n_0)}{\di \sqrt{h^2 - c^2}},
\quad   \alpha_0 = \di \sqrt{\frac{h^2 - c^2}{\lambda + h^2}}, 
    \quad n_0 =  \di \sqrt{\frac{h^2 - a^2}{h^2 - c^2}}, \quad 
     h^2 = \beta w^2 + c^2
$$
Here,  $\alpha_0$  and $n_0$  are the argument and modulus of an
incomplete elliptic integral of the first kind $F\,(\alpha_0,  n_0)$.

The derivative from the potential $U^*(\lambda)$ with
respect to  according to (6) is
$$
   \frac{\partial U^*(\lambda)}{\partial w} = 
   -\,2 \pi G \rho_0 ac^2w\, \int \limits_{\lambda}^{\infty}
    \frac{du}{\di \sqrt{(h^2 + u)^3(a^2 + u)(c^2 + u)}} =  
   -\,\frac{4 \pi G \rho_0 ac^2}{\di (h^2 - a^2)\,\sqrt{\beta}}\,
   \left[F\,(\alpha_0, n_0) - E\,(\alpha_0, n_0)\right],
$$
where $F\,(\alpha_0,  n_0)$ and $E\,(\alpha_0,  n_0)$ are incomplete elliptic
integrals of the first and second kind, respectively. The
$\alpha_0$, $n_0$ and $h^2$ values are defined above. Setting in these
formulas $z = 0$ or $y = 0$, we obtain the expressions for
the potential $U^*(\lambda)$ and its derivative on the coordinate axes $OY$ or $OZ$, respectively. It is taken into
account that $x = 0$.

Finally, on the coordinate axis $OX$, we have
$$
  w = 0, \quad   m^2(u) = \frac{x^2}{a^2 + u}, \quad    \lambda = x^2 - a^2, 
  \quad v_1 = -\,c^2, \quad v_2 = -\,a^2 - \beta x^2
$$
In this case, expression (8) of the potential $U^*(\lambda)$ should take into account that
$$
   U_2(\lambda) =   \int \limits_{\lambda}^{\infty} \frac{du}{\di \di (c^2 + u)\,\sqrt{a^2 + u}\, \sqrt{1 + \beta\, m^2(u)}}
 =  \int \limits_{\lambda}^{\infty}  \frac{du}{\di (c^2 + u)\,\sqrt{p^2 + u}} = 
 \frac{\ln \varphi_1(\lambda)}{\di \sqrt{p^2 - c^2}}, 
$$
where
$$ 
  \varphi_1(\lambda) = \di \frac{\di \sqrt{p^2 + \lambda} + \sqrt{p^2 - c^2}}{\di \sqrt{p^2 + \lambda} - \sqrt{p^2 - c^2}}, \quad p^2 =  a^2 + \beta x^2
$$
The derivative from the potential $U^*(\lambda)$ with respect to $x$
 has the form
$$
   \frac{\partial U^*(\lambda)}{\partial x} = -\,2 \pi G \rho_0 ac^2x\,
    \int \limits_{\lambda}^{\infty} \frac{du}{\di (c^2 + u)\,\sqrt{(p^2 + u)^3}} = 
   \frac{2 \pi G \rho_0 ac^2x}{p^2 - c^2}\,
   \left[\frac{2}{\di  \sqrt{p^2 + \lambda}} - 
   \frac{\ln \varphi_1(\lambda)}{\di \sqrt{p^2 - c^2}}\right]
$$
The function $\varphi_1(\lambda)$ and parameter  $p^2$  are defined
above.

Now let us calculate the derivatives of the potential
  $U^*(\lambda)$  with respect to the coordinates in the general
case:
$$
 \frac{\partial  U^*(\lambda)}{\partial R} = -\,2\pi G \rho_0 a c^2 R\,\bar R\,(\lambda), \quad R = \{x, y, z\},
\quad \bar R\,(\lambda) = \{X\,(\lambda), Y\,(\lambda), Z\,(\lambda)\}, 
\eqno (9) $$
where
$$  
    X\,(\lambda) =  \int \limits_{\lambda}^{\infty} \frac{c^2 + u}{\di \sqrt{c^2 + u}\,\sqrt{(u - v_1)^3(u - v_2)^3}}\,du =
    S_0X_0(\lambda) + S_1(\lambda),
 $$
 $$
    Y\,(\lambda) = Z\,(\lambda) =  \int \limits_{\lambda}^{\infty} \frac{a^2 + u}{\di \sqrt{c^2 + u}\,\sqrt{(u - v_1)^3(u - v_2)^3}}\,du =  S_0W_0(\lambda) + S_1(\lambda),
 $$
Here,
$$
     S_0 =  \frac{2}{(v_1 - v_2)^2\sqrt{-c^2 - v_2}}, \quad  S_1(\lambda) =  \frac{2}{(v_1 - v_2)(c^2 + v_1)} 
  \di \sqrt{\frac{\lambda + c^2}{(\lambda - v_1)(\lambda - v_2)}},
$$ 
$$
   X_0(\lambda) =  2(c^2 + v_2) E\,(\alpha, n) - (2c^2 + v_1 + v_2)\, F\,(\alpha, n),
$$
$$
   W_0(\lambda) =   \frac{(a^2 + c^2)(v_1 + v_2) + 2(a^2c^2 + v_1v_2)}{c^2 + v_1}\, E\,(\alpha, n) - 
  (2a^2 + v_1 + v_2)\, F\,(\alpha, n), 
$$
where the roots $v_1$ and  $v_2$, as well as the argument  $\alpha$
and the modulus $n$ of the elliptic integrals of the first
and second kind $F\,(\alpha, n)$ and $E\,(\alpha, n)$ are given above.

Further, it can be established that the potential $U^*(\lambda)$
 defined by formulas (7) or (8) has all the characteristic properties of a force function:
 
(1) it is a continuous function of coordinates $x, y, z$
throughout the space;

(2) it has continuous first partial derivatives
throughout the space; these derivatives have no discontinuity at the boundary of the ellipsoid, which 
follows from expression (7), in which we should set $\lambda = 0$
to obtain the internal potential;

(3) it turns to zero at infinity along with its first partial derivatives;

(4) it satisfies the Laplace equation outside the
LIEG gravitating body, and satisfies the Poisson equation inside the gravitating body ($\lambda = 0$).

The proof of these properties follows, inter alia,
from the characteristics of the  profile $\rho\,(m)$, which is a
positive, finite, continuous function and has continuous derivatives of the first and second orders. 
$\delta\,(m^2(u))$ is also such a function. Therefore, the improper integrals in formula (7) converge, and the potential 
$U^*(\lambda)$  and its first derivatives with respect to coordinates are
finite, continuous functions of their arguments. In
addition, if the PGB recedes to infinity,  also tends to
infinity. Therefore, the force function  $U^*(\lambda)$  and its
first partial derivatives vanish at infinity. The satisfaction of the Laplace and Poisson equations is verified
by calculating the partial second-order derivatives of
the potential $U^*(\lambda)$. The proof of this is omitted for
brevity.

\vskip 4 ex

4. DARK MATTER POTENTIAL OF A LAYERED 
INHOMOGENEOUS ELLIPTICAL GALAXY
\vskip 2 ex

For brevity and convenience, we will consider only
option 2 of Model 5, from which the results of option1
are obtained by elementary substitution. In this case,
the external potential of such a galaxy will be deter-
mined by the equality  $\widetilde U\,(\lambda) = U^*(\lambda) +  U_G(\lambda)$.  Here,
 $U^*(\lambda)$ represents the BM potential with the  profile $\rho\,(m)$ from (5) and is determined by equality (8), and
 $ U_G(\lambda)$  is the DM potential with an analogue of the NFW $\rho_G(m)$ profile [4, 15]:
$$
 \rho_G(m) = \frac{K}{\xi m (1 + \xi m)^2}, \quad   \xi = \di \frac{\sqrt[3]{a c^2}}{r_s}, 
\eqno (10) $$
where $r_s$ is the radius-scale of the galaxy, and the functions   $m^2(u)$  and $\Delta\,(u)$ are 
determined by equality (7).  Therefore,
$$
     U_G(\lambda) =   \pi G  a c^2 \int \limits_{\lambda}^{\infty}
      \frac{ \chi\,(u)}{\Delta\,(u)}\, du, \quad  
        \frac{\partial  U_G(\lambda)}{\partial R} = 
  -\,2 \pi G a c^2R\,\int \limits_{\lambda}^{\infty}  \frac{\rho_G(m\,(u))\, du}{(A^2 + u)\,\Delta\,(u)} 
\eqno (11) $$
Here, the  values $A^2, \ \lambda$ and function $\Delta\,(u)$ are defined
above (see Section 3), and $\chi\,(u)$ is
 $$
  \chi\,(u) =   \int \limits_{m^2(u)}^1  \rho_G(v) \, dv^2 =  
  \frac{2K}{\xi^2}\left[\frac{1}{1 +  \xi m\,(u)} - \frac{1}{1 +  \xi}\right],
 $$
So the potential $U_G(\lambda)$ will take the form
$$
      U_G(\lambda) =   \frac{2\pi G K a c^2}{\xi^2}\, \left[J_1(\lambda) + J_2(\lambda)\right], 
\eqno (12)  $$
where
$$
   J_1(\lambda) = -\,\frac{1}{1 + \xi}
   \int \limits_{\lambda}^{\infty}\frac{du}{\Delta\,(u)}, \qquad 
   J_2(\lambda) = \int \limits_{\lambda}^{\infty} \frac{du}{\left[ 1 + \xi m\,(u) \right] \Delta\,(u)}
$$
After integration, we obtain
 $$  
   J_1(\lambda)  = \di  -\,\frac{\ln \varphi\,(\lambda)}{(1 + \xi) \sqrt{a^2 - c^2}},  
   \qquad  \varphi\,(\lambda) =  \frac{\sqrt{a^2 + \lambda} + \sqrt{a^2 - c^2}}{\sqrt{a^2 + \lambda} - \sqrt{a^2 - c^2}}
\eqno (13)  $$
$$
   J_2(\lambda) =  \frac{1}{u_1 - u_2} \sum_{k = 1}^2 
 (- 1)^{k+1} \left[ \sqrt{a^2 + u_k} \,\ln \psi_k(\lambda) - \xi\, \sqrt{x^2 + w^2}\,
 \frac{\di \sqrt{f\,(u_k)}}{c^2 + u_k}\, \ln \eta_k(\lambda)\right], 
\eqno (14)  $$
where
$$
  u_1 = \frac{-\,p_2 + \sqrt{\delta_2}}{2} < 0, \quad   
  u_2 = \frac{-\,p_2 - \sqrt{\delta_2}}{2} < 0, \quad   \delta_2 = p_2^2 - 4q_2, \quad  (0 >u_1 > u_2),
$$ 
$$
 p_2 =  a^2 + c^2 - \xi^2(x^2 + w^2), \quad  q_2 = a^2c^2\left[1 - \xi^2 m^2(0)\right], 
  \quad u_0 = \frac{c^2 x^2 + a^2w^2}{x^2 + w^2},   \quad (c^2 < u_0 < a^2), 
$$
$$ 
   \eta_k(\lambda) = 
   \di \frac{2\sqrt{f\,(u_k)\,f\,(\lambda)}  + (c^2 + u_k)(u_0 + \lambda) + (c^2 + \lambda)(u_0 + u_k)} 
    {(\lambda - u_k) \left[2\sqrt{f\,(u_k)} +  2u_k + c^2 + u_0\right]}
$$
$$
   f\,(u) = (u_0 + u)(c^2 + u), \qquad \psi_k(\lambda) = \di \frac{\sqrt{a^2 + \lambda} + \sqrt{a^2 + u_k}}{\sqrt{a^2 + \lambda} -  \sqrt{a^2 + u_k}},  \quad (k = 1, 2),
$$
In the plane $OZY$, we have 
$$
   x = 0, \quad   m^2(u) = \frac{w^2}{c^2 + u},  \quad     \lambda = w^2 -  c^2 > 0 
$$
In this case, the potential $U_G(\lambda)$ is expressed by equality (12), in which the function $J_1(\lambda)$ is 
exactly the same, but the function $J_2(\lambda)$ is defined differently:
$$
  J_2(\lambda) = J_{21}(\lambda) -  \xi\, w J_{22}(\lambda),
\eqno (15) $$
where 
$$
J_{21}(\lambda) = \int \limits_{\lambda}^{\infty} 
    \frac{du}{(u + h)\,\sqrt{a^2 + u}} =  \di \frac{1}{\sqrt{a^2 - h}}\, 
     \ln \frac{\sqrt{a^2 + \lambda} + \sqrt{a^2 - h}}{\sqrt{a^2 + \lambda} - \sqrt{a^2 - h}}, \quad h = c^2 - \widetilde \xi^2w^2, \ (h < c^2 < a^2)
$$
$$
   J_{22}(\lambda) = \int \limits_{\lambda}^{\infty} 
    \frac{du}{(u + h)\,\sqrt{(a^2 + u)(c^2 + u)}} = 
   \di \frac{1}{\sqrt{(a^2 - h)(c^2 - h)}}\,\times 
 $$
 $$ 
   \times\, \left[  \ln \frac{\left(\sqrt{(a^2 + \lambda)(c^2 - h)} + 
   \sqrt{(c^2 + \lambda)(a^2 - h)}\right)^2}{\lambda + h} - 
   \ln \left(\sqrt{a^2 - h} + \sqrt{c^2 - h} \right)^2\right]
$$

Setting $w = y$ or   $w = z$  in the functions $J_1(\lambda)$  and $J_2(\lambda)$, 
we obtain the expressions for the potential $U_G(\lambda)$  on the coordinate axes $OY$ or $OZ$, respectively.

On the coordinate axis $OX$, we have
$$
   w = 0, \quad   m^2(u) = \frac{x^2}{a^2 + u},  \quad  \lambda = x^2 - a^2 > 0 
$$
For the function $J_2(\lambda)$, we then find 
$$
  J_2(\lambda) = \frac{1}{a^2 - c^2 - \xi^2 x^2}\,
  \left[\sqrt{a^2 - c^2}\, \ln \varphi\,(\lambda) - \xi\, \sqrt{x^2}\,  \ln \frac{\left(\xi\,  \sqrt{x^2} +
   \sqrt{a^2 + \lambda}\right)^2}{\lambda + c^2}\right]
\eqno (16) $$
Here, function $\varphi\,(\lambda)$ is defined above. The potential
$U_G(\lambda)$, in this case, is expressed by equality (12), in
which the expression for the function $J_1(\lambda)$ remains
exactly the same, and $J_2(\lambda)$ is defined by equality (16).

Finally, the value of the potential at the origin (in
the center of the galaxy) is
$$
  U_G^0 =   U_G(\lambda = 0) =  \frac{2\pi G K a c^2}{\xi\,(1 +  \xi)}\,
  \frac{\ln \varphi\,(\lambda = 0)}{\sqrt{a^2 - c^2}},
\eqno (17) $$

It should be noted that according to option 1 of
Model 5, the EG is considered as an inhomogeneous
elongated spheroid consisting of the BM and DM with
the corresponding $\rho\,(m)$ and $\rho_G(m)$  profiles and the
halo. If we assume that the LIEG with the halo is
bounded by an elongated spheroidal surface with
semiaxes $\widetilde a > \widetilde b = \widetilde c$, in expressions (8) and (12) for the
potentials $U^*(\lambda)$ and $U_G(\lambda)$  and their derivatives, $a$, $c$ 
should be replaced with $\widetilde a$, $\widetilde c$. Further, we will obtain
an explicit expression for the general potential $\widetilde U\,(\lambda)= U^*(\lambda) + U_G(\lambda)$ 
 according to this variant of Model 5, which is not shown for brevity. In addition,
there is also no interface between the BM and DM in
this variant, i.e., there is no need to determine the
conditions for matching the potentials  $U^*(\lambda)$ and $U_G(\lambda)$.

\vskip 4 ex
 
5. STATIONARY SOLUTIONS. LIBRATION POINTS

\vskip 2 ex

To find the stationary solutions of the system of
equations (1), we set
$$
   x =  x_0 = \mbox{const}, \quad y = y_0 = \mbox{const}, \quad z =  z_0 = \mbox{const}, 
$$
This will give us a system of algebraic equations for
finding stationary solutions in the form
$$
    \frac {\partial U\,(x_0, y_0, z_0)}{\partial x} = 0, \quad 
   \frac {\partial U\,(x_0, y_0, z_0)}{\partial y} = 0, \quad 
   \frac {\partial U\,(x_0, y_0, z_0)}{\partial z} = 0
\eqno (18)  $$
Solving the system of equations (18), we will consider
option 1 with a halo and option 2 without a halo of
Model 5. The zero solution  $x_0 = 0$,  $y_0 = 0$ and  $z_0 = 0$  
of the system of equations (18) corresponds to the
central libration point, which we denote by  $L_1 = L_1(0, 0, 0)$. 
There are two libration points on axes $OX$ and $OY$, specifically, $L_2 = L_2(x_0, 0, 0)$ and 
$L_3 = L_3(-\,x_0, 0, 0)$  on axis $OX$, and $L_4 = L_4(0, y_0,  0)$ and $L_5 = L_5(0, -\,y_0, 0)$ 
 on axis $OY$.
 
Table 1 shows the coordinates (in kpc) of collinear  $L_2, L_3$
 and triangular $L_4, L_5$  libration points calculated
according to options 1 and 2 of Model 5 for three EGs:
NGC 4374, NGC 4472, and NGC 4697 considered to
be layered inhomogeneous elongated spheroids.

 \vskip 4.0ex 

{\bf Table 1.} Coordinates (kpc) of the collinear $L_2(x_0, 0, 0)$,  $L_3(-\,x_0, 0, 0)$,  and triangular 
$L_4(0, y_0, 0)$,  $L_5(0, -\,y_0, 0)$  libration points found according to options 1 and 2 of Model 5 for three EGs. The galaxies are considered to be inhomogeneous elongated spheroids with semiaxes $a > b = c$.

 \vskip 2.0ex    
 
\begin{tabular}{|c|c|c|c|c|c|}
\hline 
  \multicolumn{1}{|c|}{} & \multicolumn{2}{|c|}{} &  \multicolumn{1}{|c|}{} & \multicolumn{2}{|c|}{} \\
    \multicolumn{1}{|c|}{Elliptical} & \multicolumn{2}{|c|}{Semiaxes, kpc}   & 
    \multicolumn{1}{|c|}{Options} & \multicolumn{2}{|c|}{Libration points}  \\
    \cline{2-3}  \cline{5-6}
  
     \multicolumn{1}{|c|}{galaxies} & \multicolumn{1}{|c|}{$a$} & \multicolumn{1}{|c|}{$b = c$} &
          \multicolumn{1}{|c|}{}  & \multicolumn{1}{|c|}{$x_0$} & \multicolumn{1}{|c|}{$y_0$}  \\
\hline   
      \multicolumn{1}{|c|}{} &  \multicolumn{1}{|c|}{} & \multicolumn{1}{c|}{} 
           & \multicolumn{1}{c|}{}    & \multicolumn{1}{c|}{}    & \multicolumn{1}{c|}{}   \\
      
  \multicolumn{1}{|c|}{NGC\ 4374} &  \multicolumn{1}{|c|}{19.947} & \multicolumn{1}{c|}{17.373}   
      & \multicolumn{1}{c|}{1)}    & \multicolumn{1}{c|}{442.547}    & \multicolumn{1}{c|}{441.301}   \\

   \multicolumn{1}{|c|}{} &  \multicolumn{1}{|c|}{}   & \multicolumn{1}{c|}{} 
      & \multicolumn{1}{c|}{2)}    & \multicolumn{1}{c|}{22.543}    & \multicolumn{1}{c|}{22.189}   \\
          \cline{4-6}
         
\hline   

           \multicolumn{1}{|c|}{} &  \multicolumn{1}{|c|}{} & \multicolumn{1}{c|}{}    & \multicolumn{1}{c|}{} 
      & \multicolumn{1}{c|}{}    & \multicolumn{1}{c|}{}      \\
             
  \multicolumn{1}{|c|}{NGC\ 4472} &  \multicolumn{1}{|c|}{22.166} & \multicolumn{1}{c|}{18.437}    
      & \multicolumn{1}{c|}{1)}    & \multicolumn{1}{c|}{532.375}    & \multicolumn{1}{c|}{530.406}   \\

   \multicolumn{1}{|c|}{} &  \multicolumn{1}{|c|}{} & \multicolumn{1}{c|}{}   
      & \multicolumn{1}{c|}{2)}    & \multicolumn{1}{c|}{24.735}    & \multicolumn{1}{c|}{24.171}   \\
          \cline{4-6}
         
\hline  

   \multicolumn{1}{|c|}{} &  \multicolumn{1}{|c|}{} & \multicolumn{1}{c|}{}    & \multicolumn{1}{c|}{} 
      & \multicolumn{1}{c|}{}    & \multicolumn{1}{c|}{}     \\ 
         
  \multicolumn{1}{|c|}{NGC\ 4697} &  \multicolumn{1}{|c|}{9.991} & \multicolumn{1}{c|}{6.304}   
      & \multicolumn{1}{c|}{1)}    & \multicolumn{1}{c|}{508.274}    & \multicolumn{1}{c|}{506.522}   \\

   \multicolumn{1}{|c|}{} &  \multicolumn{1}{|c|}{} & \multicolumn{1}{c|}{}   
      & \multicolumn{1}{c|}{2)}    & \multicolumn{1}{c|}{11.121}    & \multicolumn{1}{c|}{10.583}   \\
     \cline{4-6}
        
\hline   
 
 \end{tabular}
 
\vskip 4 ex

6. TYPE AND STABILITY OF SINGULAR POINTS

\vskip 2 ex

Singular points of the family $U = C$ are the point
at which it is impossible to construct a single tangen
plane. To determine such points, we obtain a system
ofalgebraic equations that exactly coincides with system (18) for determining libration points. Therefore
libration points $L_n$ are singular points.

To study the type and establish the stability of singular points (libration points), the force function  $U$
expanded into a Taylor series in the neighborhood of $L_n, \ (n = 1,2 \cdots 5)$, and a family of 
zero-velocity surfaces $U = C$ is written (see Section 2). Further, the
motion of the PGB near these points, which is
expressed by a system of differential equations in vari
ations, is considered. After that, the characteristi
equation of this system is written and, depending o
the roots of this equation, the stability of the libratio
points  in the sense of Lyapunov in the first approx
imation (or in the linear setting) is established accord
ing to the well-known Lyapunov theorem. The entir
procedure is described in detail by the author [2, 3]
Thus, for brevity, it is not presented here.

The type and stability of the central libration point $L_1(0, 0, 0 )$
 are the same in all models: it is an isolated
singular point, stable in the sense of Lyapunov in the
first approximation and in a nonlinear setting. Collinear libration points $L_2$ and $L_3$  found according to
Models 3, 4, and 5 in this study are conical singular
points with a cone axis $OX$ and are unstable in the
sense of Lyapunov in the first approximation, while
the triangular libration points $L_4$ and $L_5$  are singular
points with a cone axis  and are stable. Therefore,
if the PGB (e.g., a star) is very close to the triangular
libration points $L_4$ or $L_5$, it will remain there forever,
i.e., there is Hill stability.

It should be noted that the studies [16–18] also
showed the instability of libration points $L_2$ and $L_3$,  and
stability of  $L_4$ and $L_5$   in the sense of Lyapunov. In addition, nonlinear analysis showed that  
and  are stable for most of the initial conditions in the sense of
Lebesgue’s measure, excluding only some resonance
cases when instability takes place [18].

\vskip 4 ex

7. EXAMPLES OF CONSTRUCTING ZERO-VELOCITY SURFACES

\vskip 2 ex

The procedure and method for constructing zero-
velocity surfaces, or Hill surfaces, are described in
detail in the author’s paper [3]. Therefore, we will not
dwell on them here. As an example, we take the ellip-
tical galaxies NGC 4374 of the E1 type, NGC 4472 of
the E2 type, and NGC 4697 of the E4 type, which we
assume to be layered inhomogeneous elongated spher-
oids with semiaxes $a > b = c$. Below are the values of
the key parameters of these galaxies: stellar mass $M^*$
and halo mass $M_h$ (in solar masses), radius scale $r_s$ in
kpc, angular velocity of rotation of galaxies $\Omega$ in radians per million 
years, parameters $\beta$ and $K$ (in solar
masses per cubic parsec) calculated by the well-known
formula, as well as the values of the semiaxes $a, b = c$ 
in kpc and density $\rho_0$  at the center of the galaxy,
expressed in solar masses per cubic parsec:
$$
\leqno   NGC \ 4374: \  M^* = 3.38844 \cdot 10^{11} M_{\odot}, \  M_h = 1.5674 \cdot 10^{13}M_{\odot}, 
    \quad  \Omega = 0.01386, \quad  \beta = 1815, 
 $$
  $$
   r_s = 168.80, \quad  a =  19.947, \quad  b = c = 17.373, \quad  K = 0.167 \cdot 10^{-3}, \quad \rho_0 = 132.71,
$$
$$
\leqno  NGC \ 4472:\  M^* = 4.67735 \cdot 10^{11} M_{\odot}, \  M_h = 2.172 \cdot 10^{13}M_{\odot}, 
    \quad  \Omega = 0.0139, \quad  \beta = 858, 
$$
$$
 r_s = 197.0, \quad  a =  22.166, \quad  b = c = 18.437, \quad  K = 0.1467 \cdot 10^{-3}, \quad \rho_0 = 50.844,
 $$
$$
\leqno  NGC \ 4697:\  M^* = 1.4125 \cdot 10^{11} M_{\odot}, \  M_h = 6.494 \cdot 10^{12}M_{\odot}, 
    \quad  \Omega = 0.02524, \quad  \beta = 650, 
$$
$$
 r_s = 130.5, \quad  a =  9.991, \quad  b = c = 6.304, \quad  K = 0.3459 \cdot 10^{-3}, \quad \rho_0 = 194.589
 $$
 
 Figure 1 shows the constructed zero-velocity surfaces with collinear $L_1, L_2, L_3$ and triangular  
libration points  $L_4, L_5$ for EG NGC 4472 according to
Model5 in the plane $x = 0$. On the left is option 2
without a halo, and on the right is option 1 with the
galactic halo. Coordinates are given in kiloparsecs.
 
\vskip 4 ex

\begin{figure}[h]
\includegraphics[width=0.47\textwidth]{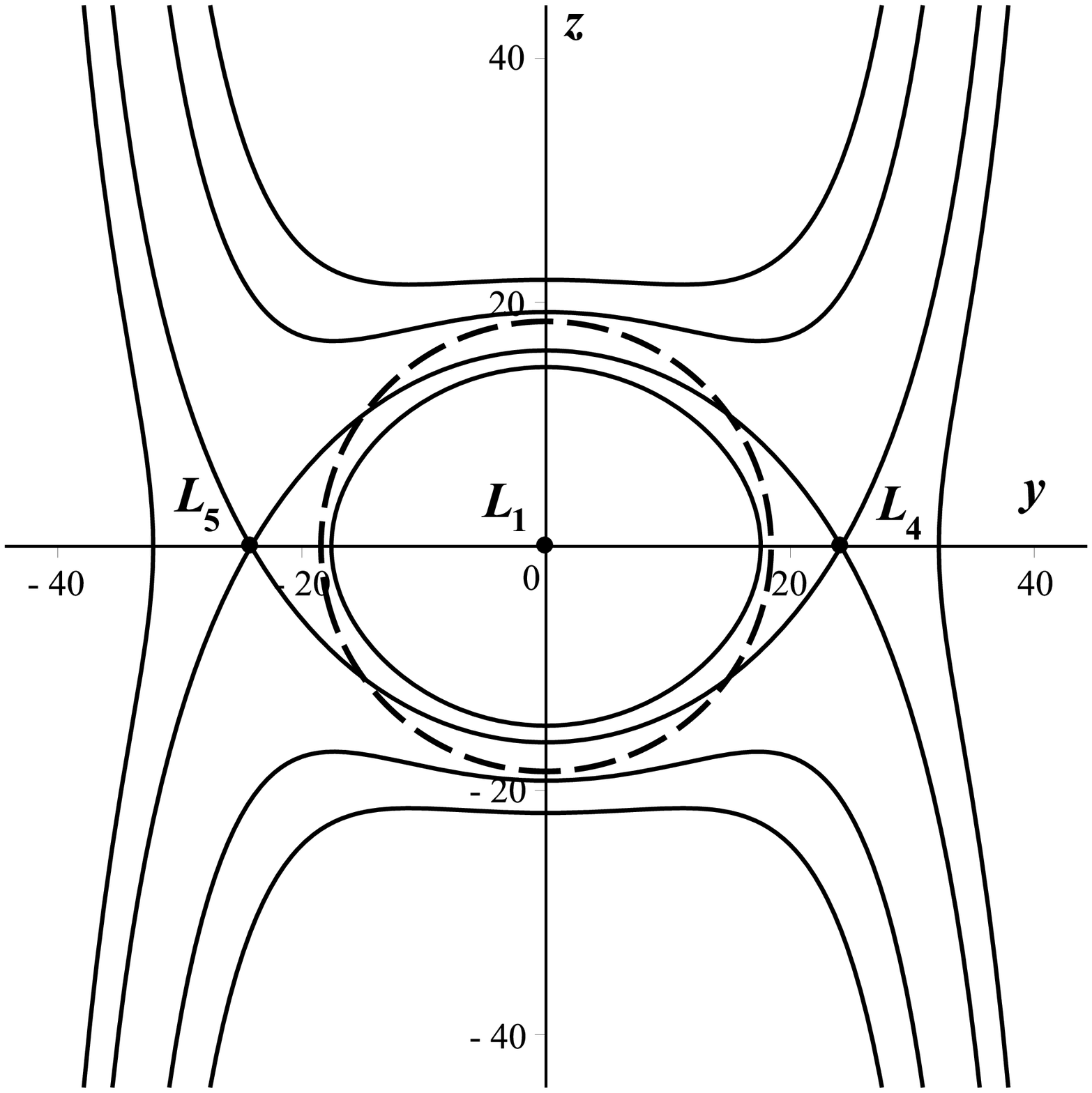}
\includegraphics[width=0.47\textwidth]{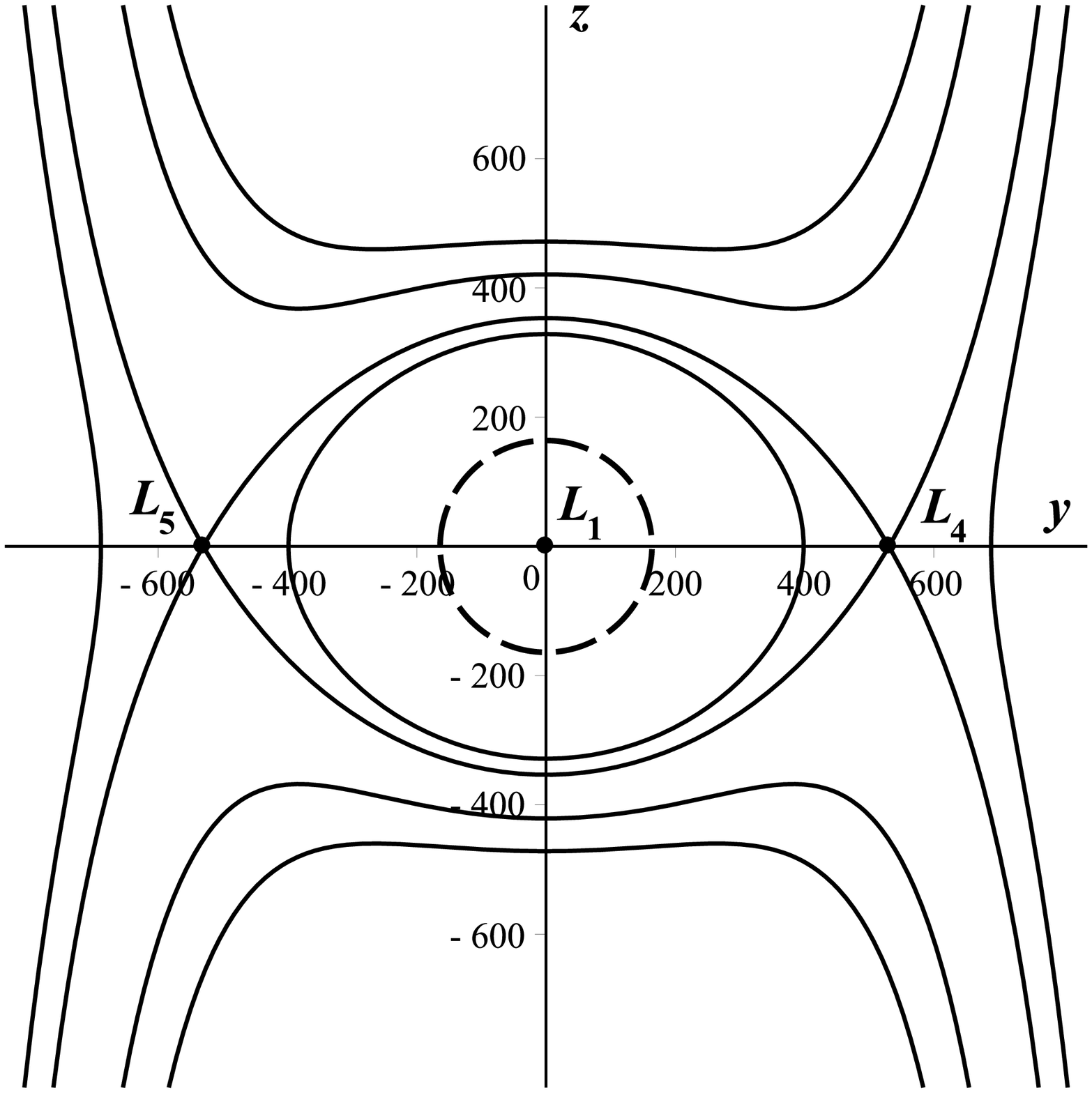}
\hfill 
\caption{Zero-velocity surfaces with collinear  $L_1, L_2, L_3$ and triangular, $L_4, L_5$  libration points for EG NGC 4472 for Model 5 in the plane $x = 0$. On the left is option 2 without a halo; on the right is option 1 with the halo. The dashed line marks the boundariesof the luminous part of the galaxy (left) and the galactic halo (right). Coordinates are given in kiloparsecs.}

 \end{figure} 
  
\vskip 4 ex

8. CONCLUSIONS

\vskip 2 ex

The spatial motion of a PGB in the gravitational
field of a LIEG is considered on the basis of the new
Model 5 for solving problems of celestial mechanics
and astrophysics. According to Model 5, an EG with a
halo (option 1) or without a halo (option 2) is a layered
inhomogeneous elongated spheroid consisting of BM
and DM. The choice of an elongated spheroid as a
model for a triaxial EG is explained by the fact that its
dynamic properties are very close to those of a triaxial
ellipsoid. In this model, there is no interface between
the BM and DM. Therefore, the determination of the
conditions for matching the potentials is not con-
sidered.

The so-called “astrophysical law” was taken as the
BM profile; it is based on the Hubble surface brightness distribution law, which adequately models the
density distribution in an EG. For the DM, an analogue of the NFW profile was taken.

An analogue of the Jacobi integral was found, the
region of the possible motion of the PGB was determined, and the zero-velocity surfaces were constructed. 
The stationary solutions (libration points)
were found to be stable in the sense of Lyapunov. Collinear libration points $L_2$ and $L_3$  found according to
Models 3, 4, and 5 are conical singular points with a
cone axis $OX$  and are unstable in the sense of
Lyapunov in the first approximation. Triangular libration points $L_4$ and $L_5$ are singular points with a cone
axis $OZ$ and are stable. The surface around the luminous part of the EG, within which the motions of the
stars or the center of mass of the GC are Hill-stable,
was determined.

The equilibrium and stability of the considered
dynamical systems according to these two models will
be studied by the author separately.

\vskip 4 ex
ACKNOWLEDGMENTS

\vskip 2 ex

The author is grateful to Prof. B.P. Kondrat’ev for valu-
able advice and comments.

\vskip 2 ex

CONFLICT OF INTEREST

\vskip 2 ex

The author declares that he has no conflicts of interest.

\vskip 4 ex

REFERENCES

\vskip 2 ex

1.S. A. Gasanov, Astron. Rep. 56, 469 (2012).

2.S. A. Gasanov, Astron. Rep. 58, 167 (2014).

3.S. A. Gasanov, Astron. Rep. 59, 238 (2015).

4.S. A. Gasanov, Astron. Rep. 65, 723 (2021).

5.B. P. Kondrat’ev, Sov. Astron. 26, 279 (1982).

6.A. V. Zasov, A. S. Saburova, A. V. Khoperskov, and
S.A. Khoperskov, Phys. Usp. 60, 3 (2017).

7.G. Bertin, R. P. Saglia, and M. Stiavelli, Astrophys. J.
384, 423 (1992).

8.M. Oguri, C. E. Rusu, and E. E. Falco, Mon. Not. R.
Astron. Soc. 439, 2494 (2014).

9.G. de Vaucouleurs, A. de Vaucouleurs, H. Corwin,
R.J. Buta, G. Paturel, and P. Fouque, Third Reference
Catalouge of Bright Galaxies (Springer, New York,
1991), Vols. 2, 3.

10.G. N. Duboshin, Celestial Mechanics, Basic Problems
and Methods (Nauka, Moscow, 1968) [in Russian].

11.H. Poincare,  Lecons sur les hypotheses cosmogoniques
(Lib. Sci. A. Hermann et fils, Paris, 1911).

12.B. P. Kondrat’ev, Potential Theory. New Methods and
Problems with Solutions (Mir, Moscow, 2007) [in Rus-
sian].

13.B. P. Kondrat’ev, Cand. Sci. (Phys. Math.) Disserta-
tion (Mosc. Phys. Tech. Inst., Moscow, 1982).

14.E. Hubble, Astrophys. J. 71, 231 (1930).

15.J. F. Navarro, C. S. Frenk, and S. D. M. White, Astro-
phys. J. 490, 493 (1997).

16.Yu. V. Batrakov, Byull. ITA 6, 524 (1957).

17.V. K. Abalakin, Byull. ITA 6, 543 (1957).

18.S. G. Zhuravlev, Sov. Astron. 18, 792 (1974).

\end{document}